\documentclass{IEEEtran}
\usepackage{cite}
\usepackage{amsmath,amssymb,amsfonts}
\usepackage{algorithmic}
\usepackage{graphicx}
\usepackage{textcomp}
\def\BibTeX{{\rm B\kern-.05em{\sc i\kern-.025em b}\kern-.08em
    T\kern-.1667em\lower.7ex\hbox{E}\kern-.125emX}}
\begin{document}
\title{{\fontsize{24}{26}\selectfont{Communication\rule{29.9pc}{0.5pt}}}\break\fontsize{16}{18}\selectfont
Analysis of High Impedance Coils both in Transmission and Reception Regimes}
\author{M.~S.~M.~Mollaei, C.C. van Leeuwen, A.J.E. Raaijmakers, and C. Simovski
\thanks{M.~S.~M.~Mollaei, and C.~Simovski are with Department of Electronics and Nanoengineering, Aalto University, Espoo, Finland (e-mail: masoud.2.sharifianmazraehmollaei@aalto.fi,konstantin.simovski@aalto.fi ). }
\thanks{C.C. van Leeuwen, A.J.E Raaijmakers are with the Department of Radiology, UMC Utrecht, Utrecht, The Netherlands (e-mail: C.C.vanLeeuwen-10@umcutrecht.nl, A.Raaijmakers@umcutrecht.nl)}}

\maketitle

\begin{abstract}
Theory of a high impedance coil (HIC) -- a cable loop antenna with a modified shield -- is discussed comprehensively for both in transmitting and receiving regimes. Understanding a weakness of the previously reported HIC in transmitting regime, we suggest another HIC which is advantageous in both transmitting and receiving regimes compared to a conventional loop antenna. In contrast with a claim of previous works, only this HIC is a practical transceiver HIC. Using the perturbation approach and adding gaps to both shield and inner wire of the cable, we tune the resonance frequency to be suitable for ultra-high field (UHF) magnetic resonance imaging (MRI). Our theoretical model is verified by simulations. Designing the HIC theoretically, we have fabricated an array of three HICs operating at 298 MHz. The operation of the array has been experimentally studied in presence of different phantoms used in UHF MRI and the results compared with those obtained for a conventional array.

\end{abstract}


\section{Introduction}
\label{sec:introduction}
Phased array antennas play a key role in modern magnetic resonance imaging (MRI) to increase spatial resolution and signal to noise ratio (SNR) compared to conventional radio-frequency coils. Since in these phased arrays, the elemental antennas are packed tightly, their coupling becomes a critical issue. In the transmitting regime, this coupling results in cross-talks, inter-channel scattering and lower SNR. In the receiving regime, it results in creation of secondary electromagnetic fields over neighbouring antennas that destroys the signal and can be also treated as the decrease of SNR \cite{r1,r2,r3,r4}.

For scanning specific parts of a human body such as head, arrays of loop antennas are used. In many MRI schemes, the methods of decoupling the loop antennas imply their partial overlapping. In this way one can cancel out the mutual inductance between adjacent loops. However, in arrays with many loop antennas in a cylindrical configuration applied in ultra-high field (UHF) MRI, these techniques are not very advantageous because the operation frequency is rather high, the coupling of loops is not purely inductive and non-neighbouring loops keep noticeably coupled \cite{r3}.

Recently, a novel approach to the decoupling of loop antennas was suggested which uses the property of a Huygens element to create a null of the electromagnetic field in the near zone \cite {r10}. In this coils, a lumped capacitor is located at a top point of the loop (which is opposite to the feeding point). When this capacitance is sufficiently small, the current distribution in the coil is not uniform -- besides of a magnetic dipole mode (uniform current) an electric dipole mode (in-phase with the magnetic current on the top of coil and opposite-phase on the bottom) are excited. Properly choosing the value of this capacitance, magnetic coupling coefficient and electric coupling coefficient become balanced and opposite in phase. This grants the decoupling for two adjacent loops having nulls of their near-field pattern on the line connecting their centers. However, the coupling of non-neighboring antennas keeps noticeable. Moreover, this technique is not convenient for scanning the movable organs as hand and knee \cite{r12}.

In \cite{r12}, a new coil with high input impedance called high impedance coil (HIC) was suggested. This HIC is introduced as a receiving loop antenna made of a coaxial cable with a gap in the shield on the loop top and the central conductor is connected to the receiver through a hole in the shield on the loop bottom. The high impedance does not allow the induced current to flow that results in the absence of scattering. This antenna was designed for a high-field MRI (operating frequency 123 MHz). In this paper, we aim to modify this receiving antenna qualitatively so that to make it suitable for UHF MRI. Here not only the operation frequency is much higher
but also the antenna should operate in the transceiver regime.

The possibility of a HIC to operate in the transmission regime was not studied in \cite{r12}. In work \cite{r13}, the authors designed a similar coil operating in the transmission regime in the range of 300 MHz
and used it for scanning knee and hand in a UHF MRI setup. The main difference compared to \cite{r12} (besides of evidently different geometric parameters) was a capacitor connected to the antenna in parallel to the receiver instead of a parallel inductor used for the antenna matching in the initial work \cite{r12}. In \cite{r13}, authors used one HIC matched as in \cite{r10} for receiving the echo-signal and another HIC matched with a capacitance for transmitting the primary signal. However, is it an optimal solution -- is it really impossible to merge a receiving and transmitting HICs into one HIC operating in a transceiver regime? The authors of \cite{r13} do not answer this question and even do not prove an advantage of their transmitting HIC compared to a conventional coil used in a UHF MRI (see e.g. in \cite{r10}). In fact, the authors of \cite {r13} do not explain how these coils work: why in one case one needs a matching inductance and in another case -- a capacitance? The physics of a HIC has been cryptic and the potential of its development unclear.

In the next section, we discuss the underlying physics of a HIC in both transmission and reception regimes. We prove that in the transmission mode the HICs introduced in \cite {r12,r13} have no advantages compared to conventional coils. Further, we show that even in the receiving regime a low-impedance reverse pre-amplifier is needed to suppress the current in the coil. Finally, using the perturbation approach we design a modified HIC operating in the transceiver regime. We prove numerically and experimentally that the suggested HIC is advantageous compared to a conventional transceiver coil.

\section{Theory of HIC}

\subsection{Transmission regime}

\begin{figure}[t!]
\centering
\includegraphics[width=0.95\linewidth]{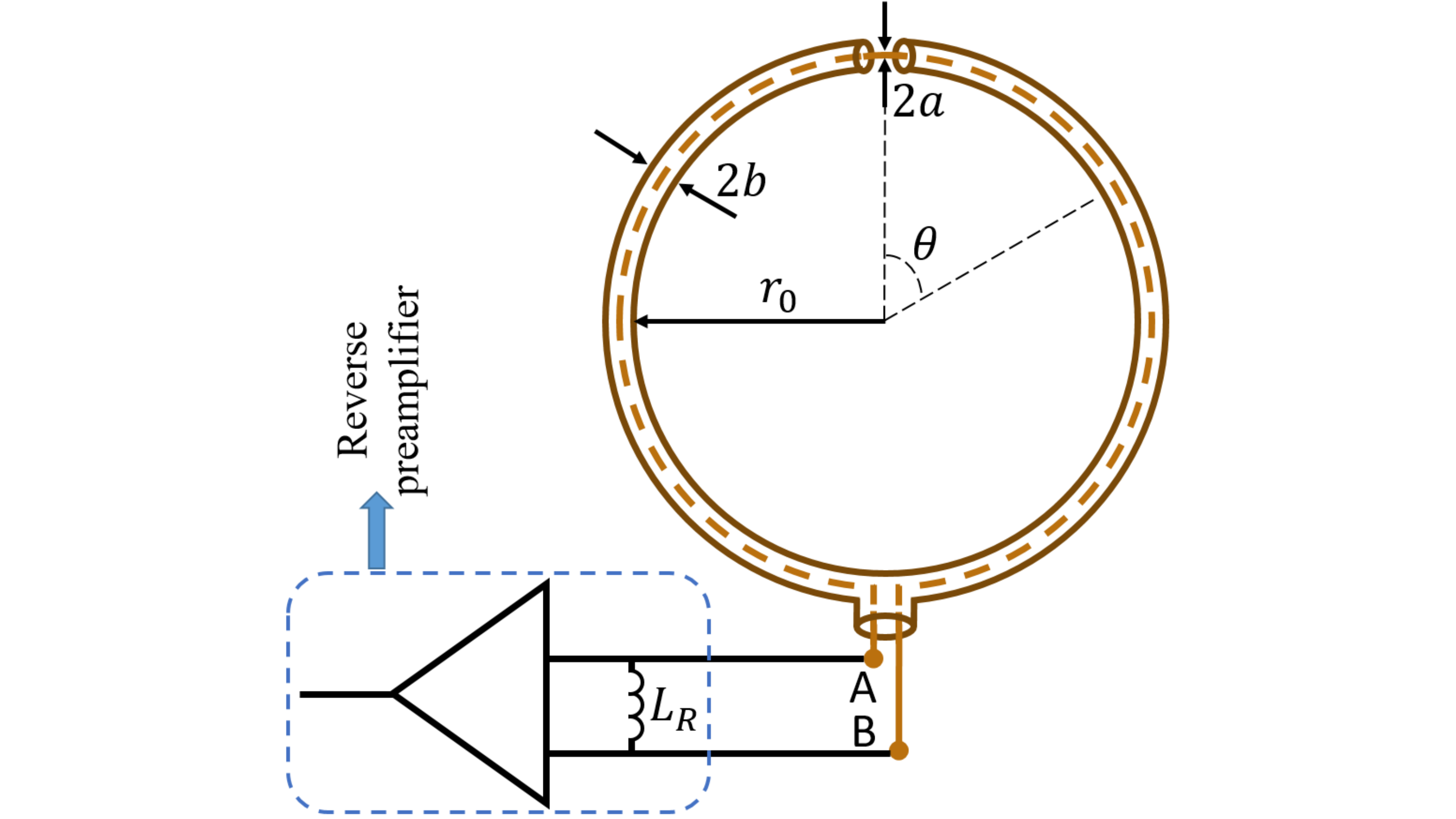}
\caption{Schematics of a HIC suggested in \cite{r12}.}
\label{f4}
\end{figure}

\begin{figure}[t!]
\centering
\includegraphics[width=0.5\linewidth]{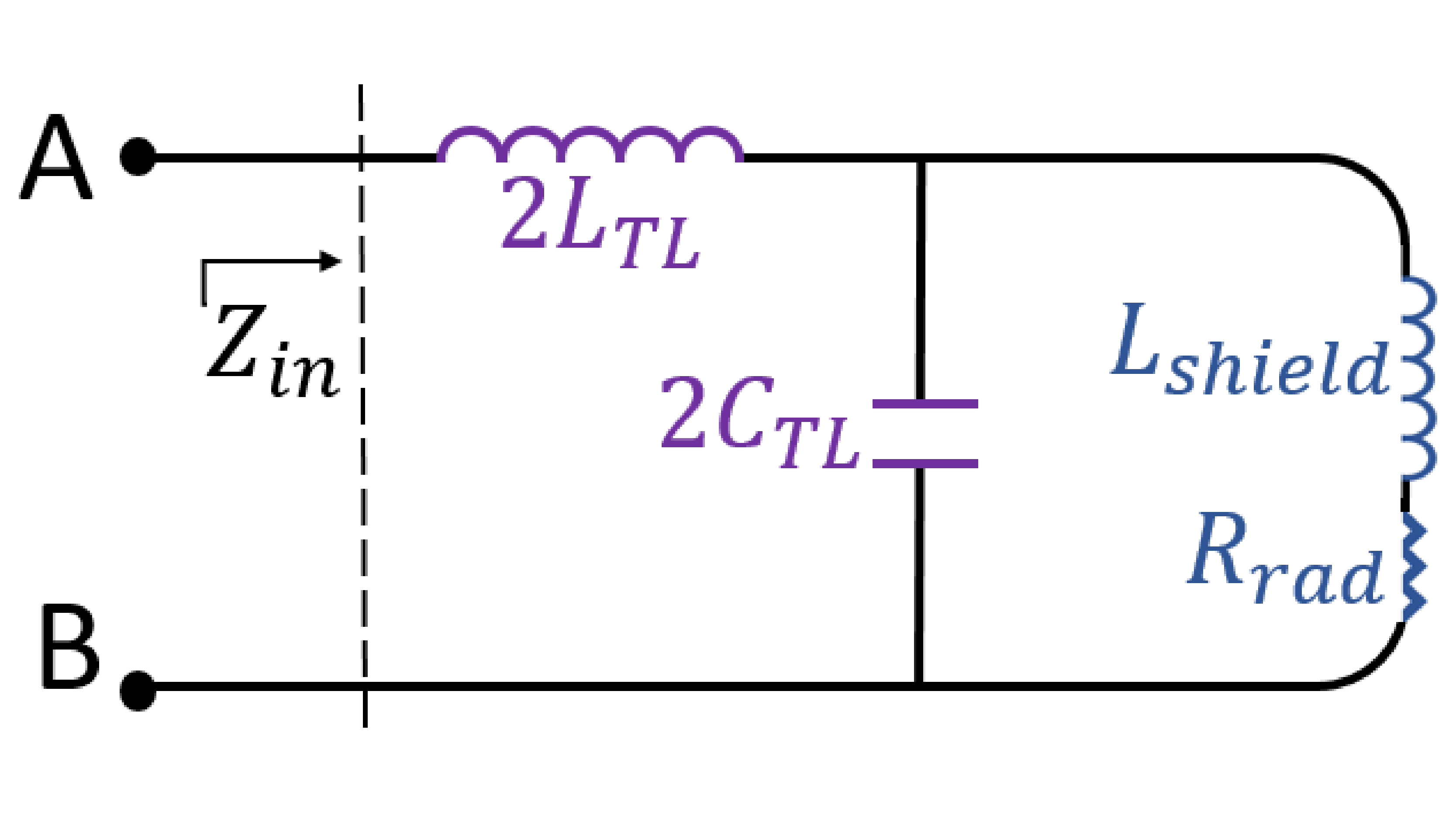}
\caption{Circuit model of a HIC from work \cite{r12} in the transmission mode.}
\label{f6}
\end{figure}

Schematic view of HIC introduced in \cite {r12} and \cite {r13} is shown in Fig.~\ref{f4}. Due to the size of the coil and the resonance frequency ($r_0=40$ mm and 123 MHz in \cite {r12} used for 3 T MRI), current distribution over outer part of the shield is uniform \cite {r11}. In accordance to the theory of loop antennas, only the magnetic mode is present. As to the input impedance, it can be found from circuit model. In this model, the antenna represents two transmission lines formed by the inner wire and the inner side of the shield connected via the load formed by the outer side of the shield. Both coaxial lines are connected to this external loop by the cut in the shield. The corresponding circuit model shown in Fig.~\ref{f6} was, in fact, suggested long ago in \cite{r11}. Here $L_{TL}$ and $C_{TL}$ are, respectively, inductance p.u.l. and capacitance p.u.l. of the coaxial cable multiplied by $\pi r_0$. $R_{rad}$ is radiation resistance and $L_{shield}$ is the inductance both referring to the outer surface of the cable loop. Radiation resistance of a circular loop of radius $r_0+b$ is calculated as that of a loop magnetic dipole \cite{r12} and the inductance is as follows \cite{r11}:
\begin{equation}
L_{shield} \approx \mu r_0 (ln \frac{8 r_0}{b}-2)
\end{equation}
With this model, we calculated the resonance frequency very close to that found in \cite{r13} from a different model. In \cite{r13} the transmission regime was described by the same
model as in \cite{r12} though the last one was obtained for the receiving regime and in \cite{r11} it was noticed that the circuit models for two regimes must be different.
To obtain the resonance frequency correctly from a wrong model is possible only for some specific values of the design parameters which were used in \cite{r13}.

\begin{figure}[t!]
\centering
\includegraphics[width=1\linewidth]{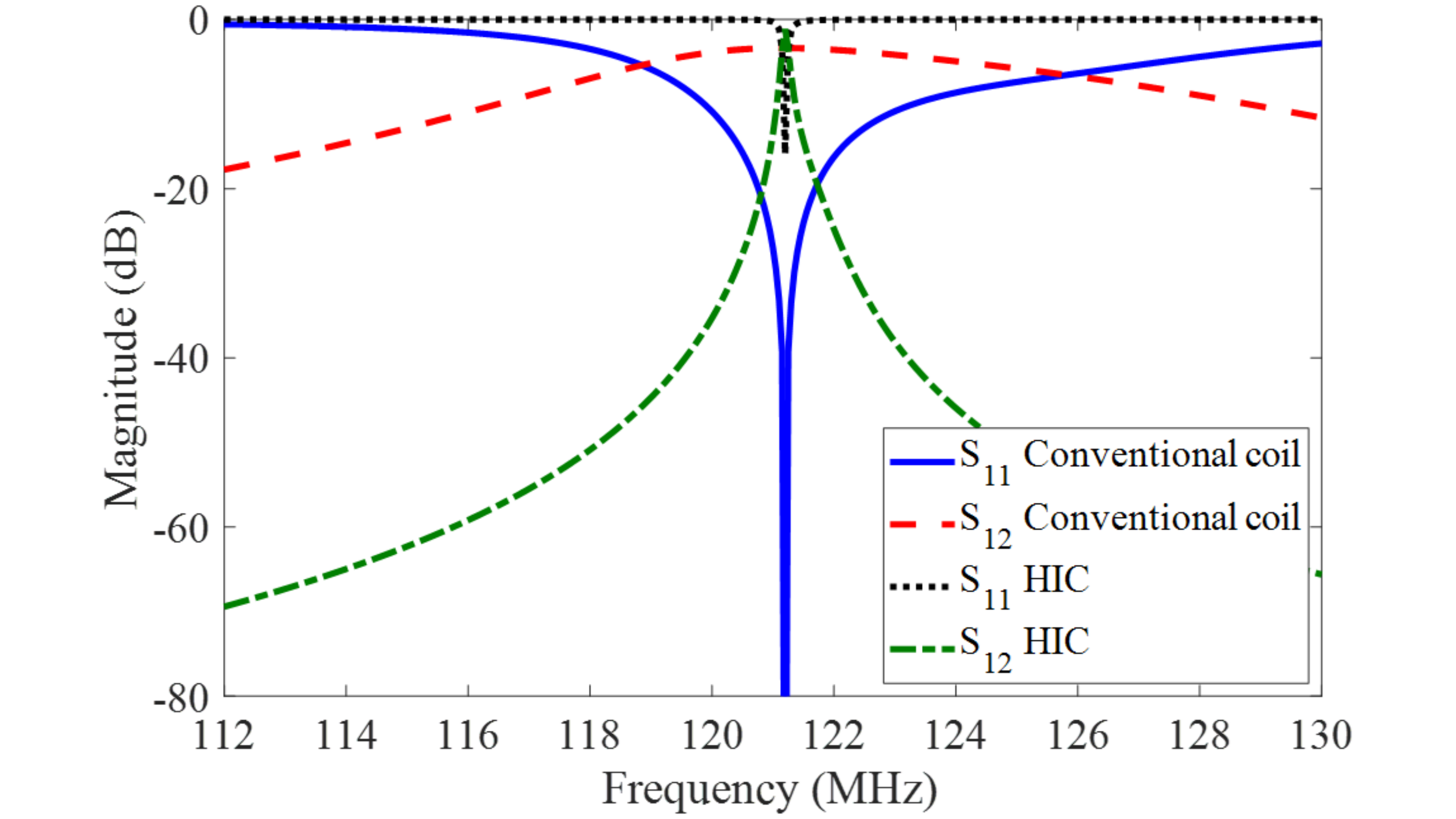}
\caption{S-parameters of two closely (the gap equals $\lambda /33$) located conventional loops and those of two closely located HICs.}
\label{f7}
\end{figure}

At the resonance frequency, the input impedance is real, and this resistance drastically exceeds $R_{rad}$. Therefore, in order to radiate the needed signal power we need a very high input voltage of the coil \cite {r13}. Thus, the transmitting HIC acts as a conventional loop antenna loaded by a capacitor on top (as it was correctly noticed in \cite {r11}). Contrary to what was claimed in \cite{r13}, though the input impedance at the resonance is high compared to $R_{rad}$, the coupling coefficient is the same and the electromotive force induced by an adjacent HIC when recalculated to the antenna input (points A and B) of a reference HIC is comparable with the voltage feeding the reference HIC. To decouple the neighbouring HICs, one still needs to overlap them as if they were usual loop antennas (whereas the non-neighbouring HICs keep not decoupled). Our simulations show that the coupling between transmitting HICs in a planar array is higher than that in a similar array of usual capacitively loaded loops of the same size. Fig.~\ref{f7} shows the simulated S-parameters for two HICs and for two usual loop antennas resonating at the same frequency when the distance between the edges of the loops is the same ($\lambda /33$) and the overall sizes of the loops are the same. We see that $S_{12}$ for a pair of HICs is equal at the resonance $-1.13$ dB, and for two conventional coils it is lower -- $S_{12}=-3.36$ dB. This is not surprising -- we have already noticed, that in the loops with capacitive loads the electric coupling and the magnetic coupling nearly cancel out. We have also simulated a structure in which two transmitting HICs (and two conventional loops for comparison) overlap so that to nullify the mutual inductance. Overlapping these HIC we have decreased the mutual coupling to the same level as for overlapping loops. Thus, we still did not see advantages of HIS. Moreover, our simulations have shown that in presence of the human body phantom, another drawback of transmitting HICs manifests. This is higher sensitivity to the phantom parameters for the input impedance of the antenna. Thus, to match these HICs will be in practice more difficult that the conventional coils. To conclude, contrary to what was claimed in \cite{r13}, the HIC suggested in \cite{r12} is not advantageous in the transmission mode.

\subsection{Receiving regime}

\begin{figure}[t!]
\centering
\includegraphics[width=0.95\linewidth]{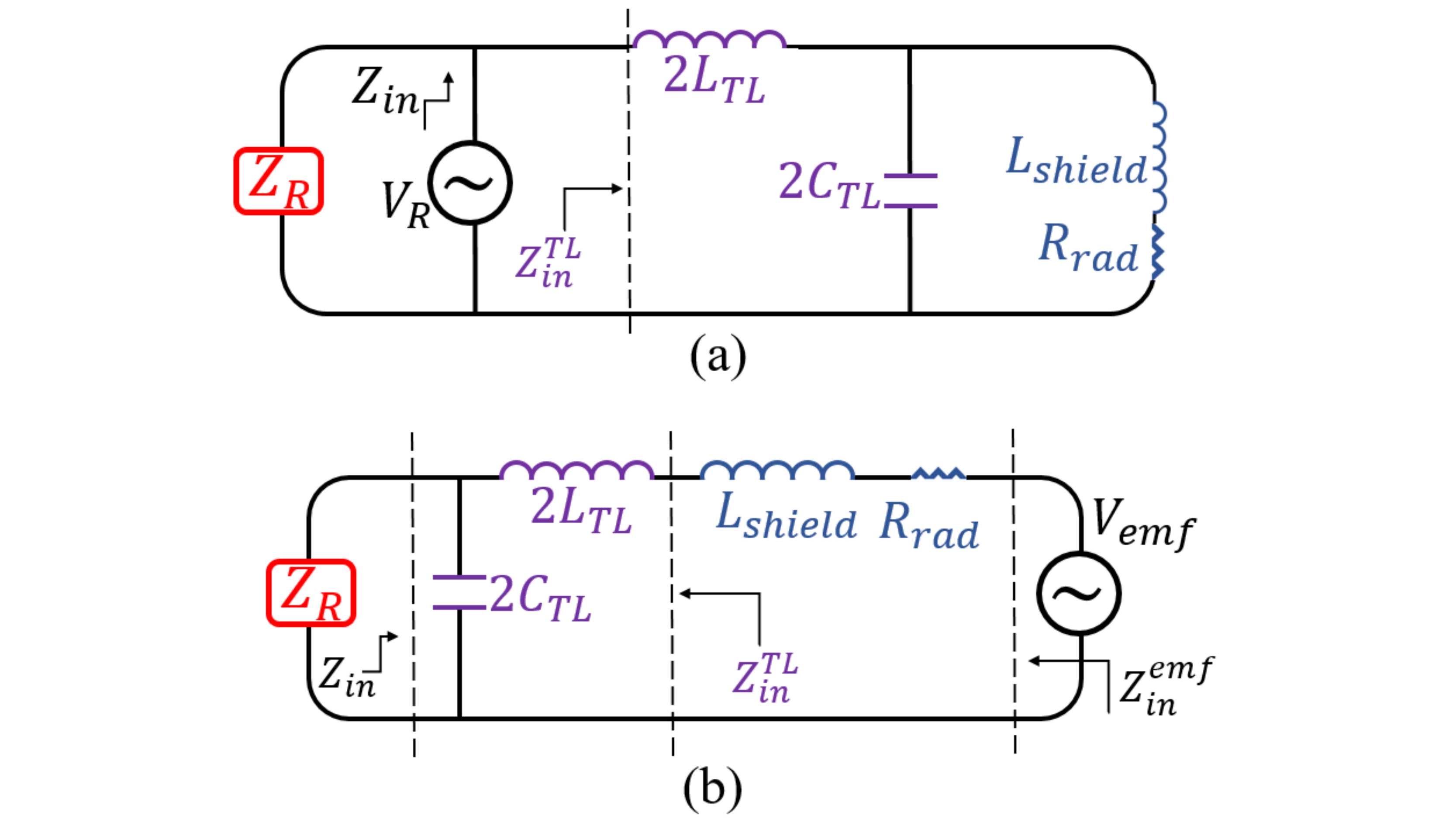}
\caption{(a) Wrong circuit model of a receiving HIC (the same as in the transmission mode). (b) Correct circuit model of the HIC in the receiving regime.}
\label{f10}
\end{figure}

As we have seen, the antenna suggested in \cite{r12} was, first, introduced in work \cite{r11}, where its correct model for the transmission regime was presented. 
As to the receiving regime, possibility of better decoupling in the antenna array compared to usual loops was mentioned in \cite{r11}. However, the receiving regime of the HIC was not studied and its peculiarities were  unexplained. Due to the lack of explanation, a wrong understanding of a receiving HIC became possible. Really, its input impedance in the receiving regime is
defined in the same place as in the transmission regime. The electromotive force (EMF) $V_{emf}$ induced in the antenna also can be recalculated to the effective voltage $V_R$ applied to the receiver in the same place.
In both transmission and reception regimes, all circuit parameters of the antenna are the same. Therefore, the equivalent circuit seems to be that shown in Fig.~\ref{f10}(a), where $Z_R$ is the output impedance of the receiver.  In this model, the small inductance $2L_{TL}$ acts nearly as a short-circuit connector, and the capacitance $2C_{TL}$ forms a parallel circuit with the inductance $L_{shield}$. Since the input resistance of a parallel circuit at the resonance is very large, the voltage induced at the input of the receiver feels the input impedance $Z_{in}^{TL}$ of a double transmission line loaded by a loop formed by the exterior of the shield. Since the voltage $V_R$ is finite, $Z_{in}\rightarrow\infty$ and the current flowing at the antenna input is negligibly small. Thus, the receiver with impedance $Z_{R}$ is excited by an ideal voltage source $V_R$ which is almost decoupled with the external loop. The receiver is driven solely by an induced voltage, induced currents are not involved. Therefore, two receivers connected to two HICs in an array should be decoupled. This is the wrong understanding of a receiving HIC in works \cite{r12,r13}. 

\begin{figure}[t!]
\centering
\includegraphics[width=1\linewidth]{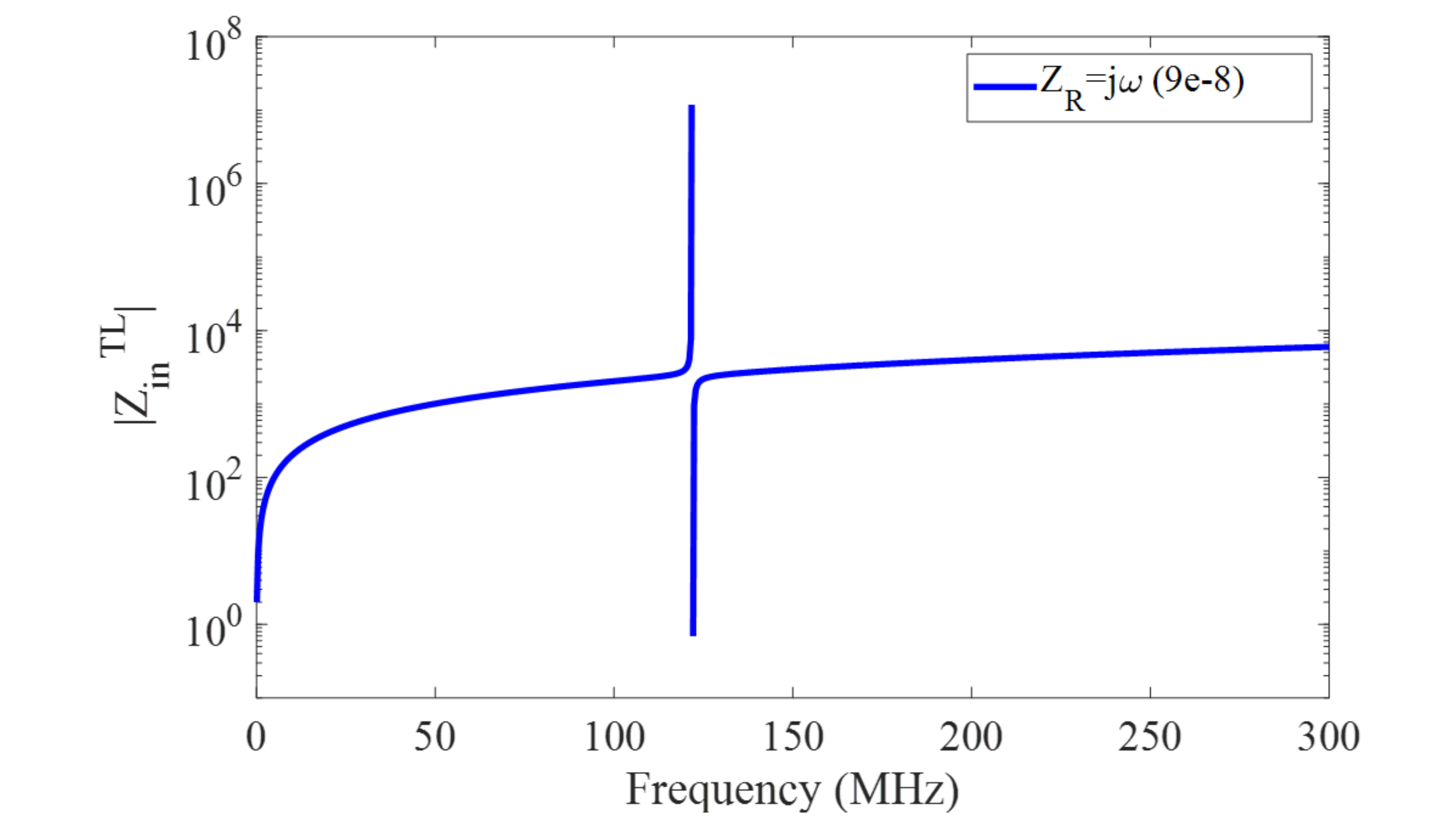}
\caption{Input impedance of the transmission line depicted in Fig.~\ref{f10}(b) versus frequency (log scale). Design parameters of a HIC are taken from \cite{r12}.}
\label{f11}
\end{figure}

In fact, the mutual coupling is not determined by the current flowing at the antenna input, whose smallness is not relevant. The effective antenna current responsible for the coupling flows in the same place where the induced EMF is formed. For a HIC $V_{emf}$ is referred to the antenna top where the receiving (and scattering) loop formed by the cable exterior is connected to two halves of the cable loop mutually connected on the bottom through the load $Z_R$. Thus, $V_{emf}$  is connected through $L_{shield}$ (responsible for $V_{emf}$) and $R_{rad}$ (responsible for scattering) to the input impedance $Z_{in}^{TL}$ of a loaded transmission line. Since two semicircles of cable are connected  by $Z_R$, this $Z_{in}^{TL}$ is different from that shown in Fig.~\ref{f10}(a) which is a wrong circuit for the receiving regime. Fig.~\ref{f10}(b) depicts the correct circuit model in the receiving regime, where we clearly distinguish the impedance $Z_{in}$ seen by the receiver and the impedances seen by the induced voltage. We see that $Z_R$ shunts the capacitance $2C_{TL}$, and a low impedance of a receiver does not grant the decoupling. In this case, our short transmission line is nearly short-cut and the antenna current flowing through the voltage source $V_{emf}$ is high. However, adding an inductance $L_R$ as it was done in \cite{r12} and shown in Fig.~\ref{f4}, we engineer a parallel circuit formed by $2C_{TL}$ and $L_R$. Since the resonant load of the transmission line to which the source is connected turns large at the operation frequency,  the relevant antenna current turns small. A preamplifier with high output, also shown in Fig.~\ref{f4} (note that its necessity was mentioned in \cite{r11} without any explanation) ensures
that the moderate output impedance of the receiver does not shunt the parallel circuit seen by the antenna current. In our model, the input impedance for the induced voltage is as follows:
\begin{equation}
{{Z_{in}^{emf}} = Z_{in}^{TL} + Z_{L}^{shield} + R_{rad}}
\end{equation}
where
\begin{equation}
{{Z_{in}^{TL}} = Z_0 \frac{Z_R + jZ_0 tan(\beta l)}{Z_0 + jZ_R tan(\beta l)}}.
\end{equation}
Here $Z_0$ is characteristic impedance of the cable, $l=2\pi r_0$ is the total length of the cable and $\beta\approx \omega/c$ is propagation constant in it.
As to $Z_R$, it is a parallel connection of $L_R$ and the pre-amplifier output impedance.
The antenna input impedance seen by the receiver is calculated as follows:
\begin{equation}
{Z_{in}}^{-1}= 2j\omega C_{TL}+{1\over R_{rad}+j\omega ({L}_{shield}+2L_{TL})}.
\end{equation}
If ${Z_{R}}$ is very small, it shorts the capacitance $2C_{TL}$, the induced voltage is loaded by a small inductance ${L}_{shield}+2L_{TL}$, the current flowing through $V_{emf}$ is large and the HICs in the array are strongly coupled. If ${Z_{R}}$ is very large, the induced voltage is loaded by a series circuit with inductance ${L}_{shield}+2L_{TL}$ and capacitance $2C_{TL}$. This circuit resonates exactly at the same frequency where $Z_{in}$ experience the parallel resonance i.e. at the operation frequency of the HIC. Then the current flowing through $V_{emf}$ is large again. Both these regimes do not allow the decoupling. Meanwhile, if ${Z_{R}}=j\omega L_R$ and $L_R$ is properly chosen, the induced voltage in a HIC sees a parallel resonance at the operation frequency of the antenna, and such HICs in an array are really decoupled as in \cite{r12}. 

In Fig.~\ref{f11} we present $Z_{in}^{TL}$ corresponding to Fig.~\ref{f10}(b) for ${Z_{R}}=  j \omega (9e-8)$. In these calculations, we used Eqs. (2) and (3) instead of a simplistic model with lumped elements as depicted in Fig.~\ref{f10}(b). The design parameters of the HIC and the value $L_{R} =  9e-8$ Hn were taken from \cite{r12} where this inductance was connected in parallel to a pre-amplifier with mega-Ohm output. We see that the inductive matching grants not only huge resonant values to $|Z_{in}^{TL}|$ but also large values beyond the resonance. This is so because our short transmission line (its $\lambda/4$ resonance occurs above $300$ MHz) is loaded by a rather large inductance $L_R\gg L_{shield}$. Therefore, the decoupling functionality of a HIC in the receiving regime is broadband. Thus, our model is in line with unexplained claims of \cite{r11} and with practical results of \cite{r12}.

\begin{figure}[t!]
\centering
\includegraphics[width=1\linewidth]{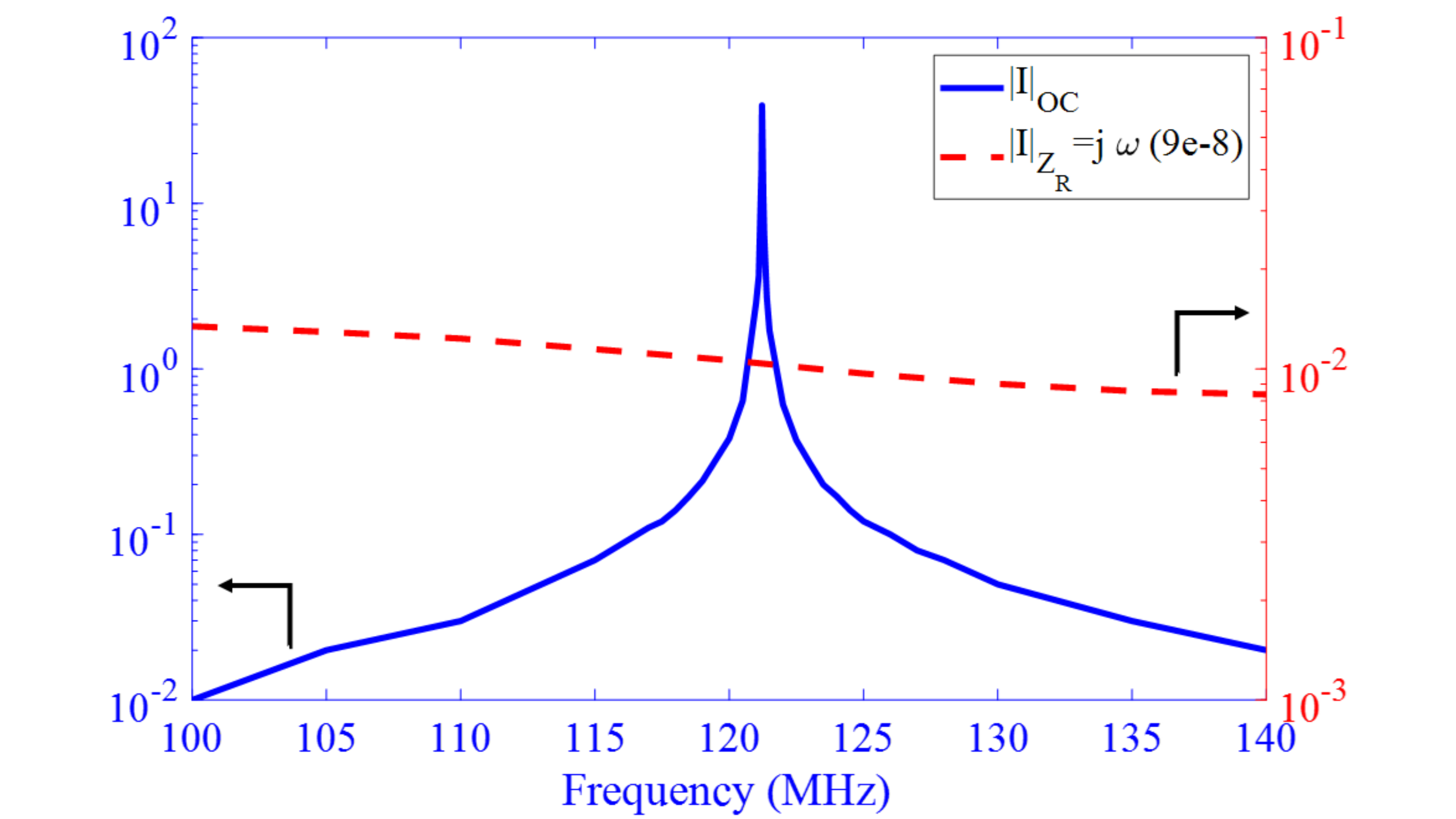}
\caption{Induced current over the HIC with different loadings.}
\label{f12}
\end{figure}

For better validation, the current induced on the exterior of a receiving HIC shied has been calculated in CST Microwave Studio. The source was a plane wave with a magnetic field orthogonal to the HIC plane. The current has been calculated for two values of $L_{R}$: $L_{R}\rightarrow \infty$ (open-circuit case) and $L_{R} =  9e-8$ that as we have seen grants the parallel resonance at the antenna operation frequency for both antenna input and output. The result is presented in Fig.~\ref{f12}. It confirms what we have claimed above: when the port is open, the induced voltage is really connected to a series circuit and this harmful resonance occurs at the antenna operation frequency. Then the induced current increases drastically i.e. in an array our HICs will be coupled strongly. When the port impedance is inductive and this inductance is sufficiently large, the harmful series resonance for the induced current does not occur and the current keeps small in the range $50-300$ MHz. Now, it is pretty clear why  the HICs were decoupled in a receiving array \cite{r12}, why the decoupling of the transmitting HICs required a matching capacitor, and why it was not better than that in an array of capacitively loaded loops.

\section{Designing Transceiver HIC}
\label{sec:guidelines}

This insight was only a way to the main task of this papery -- design of a transceiver HIC for scanning the human head in 7 Tesla MRI with 8 channel pre-amplifier. It means that the radius of the HIC should be larger than the HICs of \cite {r12,r13} where either the frequency range was lower or the scanning object was hand or knee. Namely, the radius of a transceiver HIC should be $r_0 = 80~mm$ to cover whole head with 8 coils. In this case, the perimeter of the coil is $P_{HIC} = 2 \pi r_0 \approx \lambda /2$ which means that the HICs of works \cite{r12,r13} cannot be considered as electromagnetically small in the range of 300 MHz. In these HICs, the high-order multipoles will be induced at such frequencies. This effect will make an analytical model and the design procedure very difficult. A transceiver HIC should be designed in such a way that only two fundamental modes of the coil -- magnetic dipole and/or electric dipole -- would exist at the operation frequency.  Moreover, it should be obtained in the situation when the electromagnetic size of the coils is larger than in works \cite {r12,r13}. So, we we have to raise the fundamental resonance of a coil from 123 MHz to nearly 300 MHz with a simultaneous increase of the HIC size. Next, we have to reduce $|S_{12}|$ for two neighboring HICs in the transmission regime. How to do it?

\begin{figure}[t!]
\centering
\includegraphics[width=0.9\linewidth]{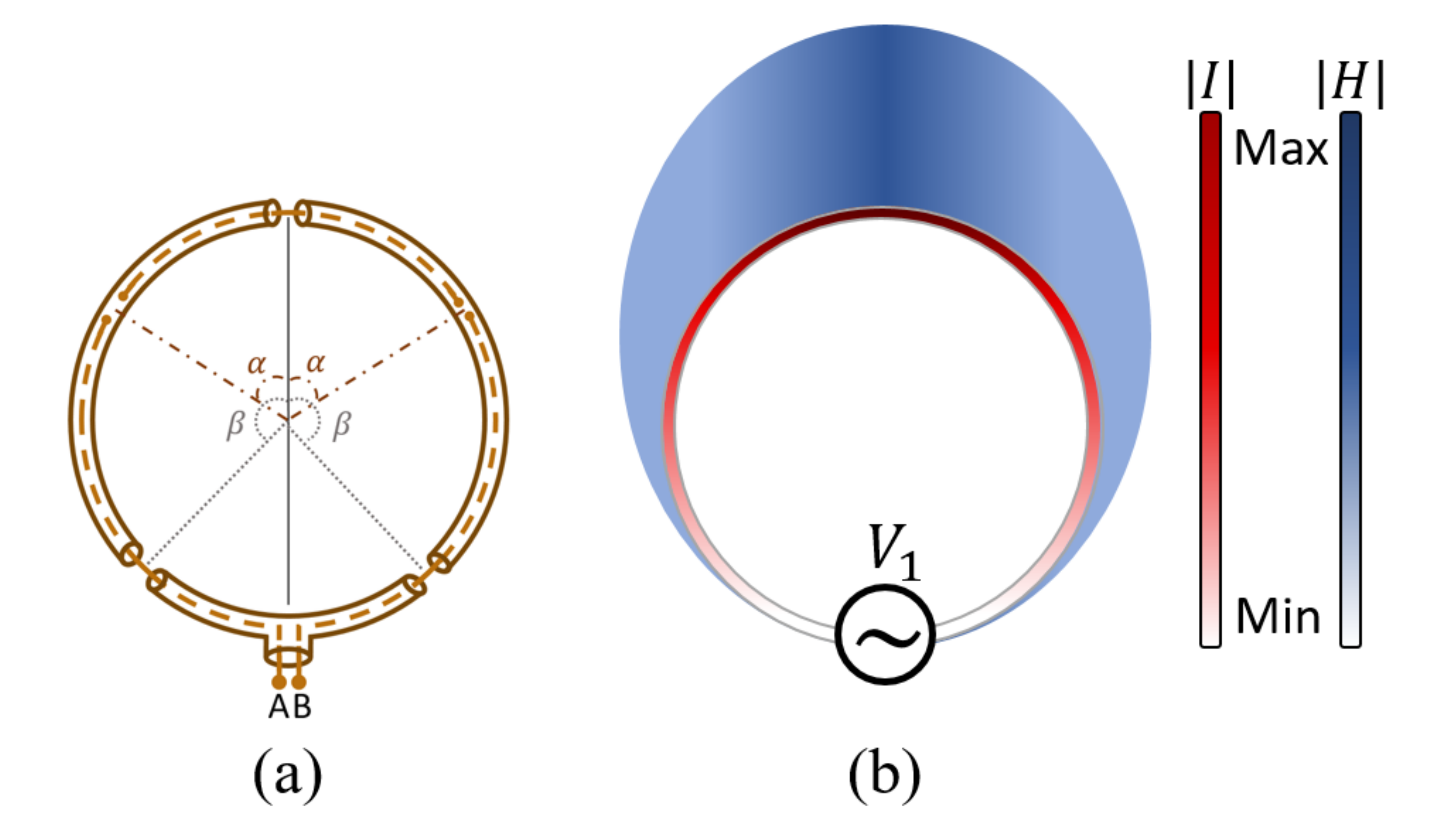}
\caption{(a) Schematic view of the new coil with added gaps in positions $\alpha$ and $\beta$. (b) Current distribution over the external loop and radiated magnetic field in the near zone (both magnetic and electric dipoles are balanced at the operation frequency 298 MHz).}
\label{f13}
\end{figure}

We have solved the second problem almost in the same way as discussed above. It is possible to engineer a Huygens regime in a HIC obtaining an electric dipole mode due to a capacitive load of the external loop. It implies additional cuts of the shield. If the optimal position of the cut is not the loop bottom (a feeding point), we should not make one cut aside the loop. Otherwise, we will make our HIC bianisotropic, bring quadrupole components in the loop current and strongly complicate our task. Thus, we should make two symmetrically located cuts in the shield. Next, according to the perturbation approach (see e.g. in \cite{harrington}), we add a perturbation at the points where the induced current component corresponding to the magneto-dipole mode is high enough. This perturbation should result in the raise of the resonance frequency for the magnetic mode. This perturbation can be a simple cut of the inner wire. These two cuts (also symmetrically located) do not break the current because two broken ends will be coupled via the interior side of the shield. Depending on the position where we add these cuts (near the maxima of the charge distribution or near the maxima of the current distribution), these cuts can be modeled as loading elements -- either inductors or capacitors.  The exact position of these gaps should be found numerically. Fig. \ref{f13}(a) shows the new HIC with cuts of the inner wire (angles $\alpha$) and the shield (angles $\beta$). These angles chosen so that our large HIC resonates at 298 MHz (frequency of the magnetic resonance of protons) with balanced electric and magnetic current modes will be specified in the next section.

For this properly designed HIC, we show in Fig. \ref{f13}(b) by color both near-zone magnetic field distribution in the loop plane and current distribution over the external loop. If two such HICs are positioned back to back (upside down), they do not produce magnetic flux over each other and are decoupled not only in the reception regime but also in the transmission regime. However, as well as for the conventional loops this arrangement cannot be extent to arrays with more than two coils. In this case we may arrange the chain of HICs with alternating locations of the antenna inputs. Then the closest coils will be weakly coupled because in Fig. \ref{f13}(b) we see that the near field aside the loop is quite weak. Below we will see below that this solution grants a reduction of the coupling in the array with respect to the known analogues.

Notice, that in the receiving regime, the performance of our coil is quite similar to that of a HIC from \cite{r12}. Of course the non-uniformity of the induced current in the case of three gaps in the cable shield 
influences the near field coupling coefficients. However, the main factor of decoupling keeps the same -- the high impedance seen by the induced voltage. 

\section{Numerical Verification}
\label{sec:numerical}

\begin{figure}[t!]
\centering
\includegraphics[width=1\linewidth]{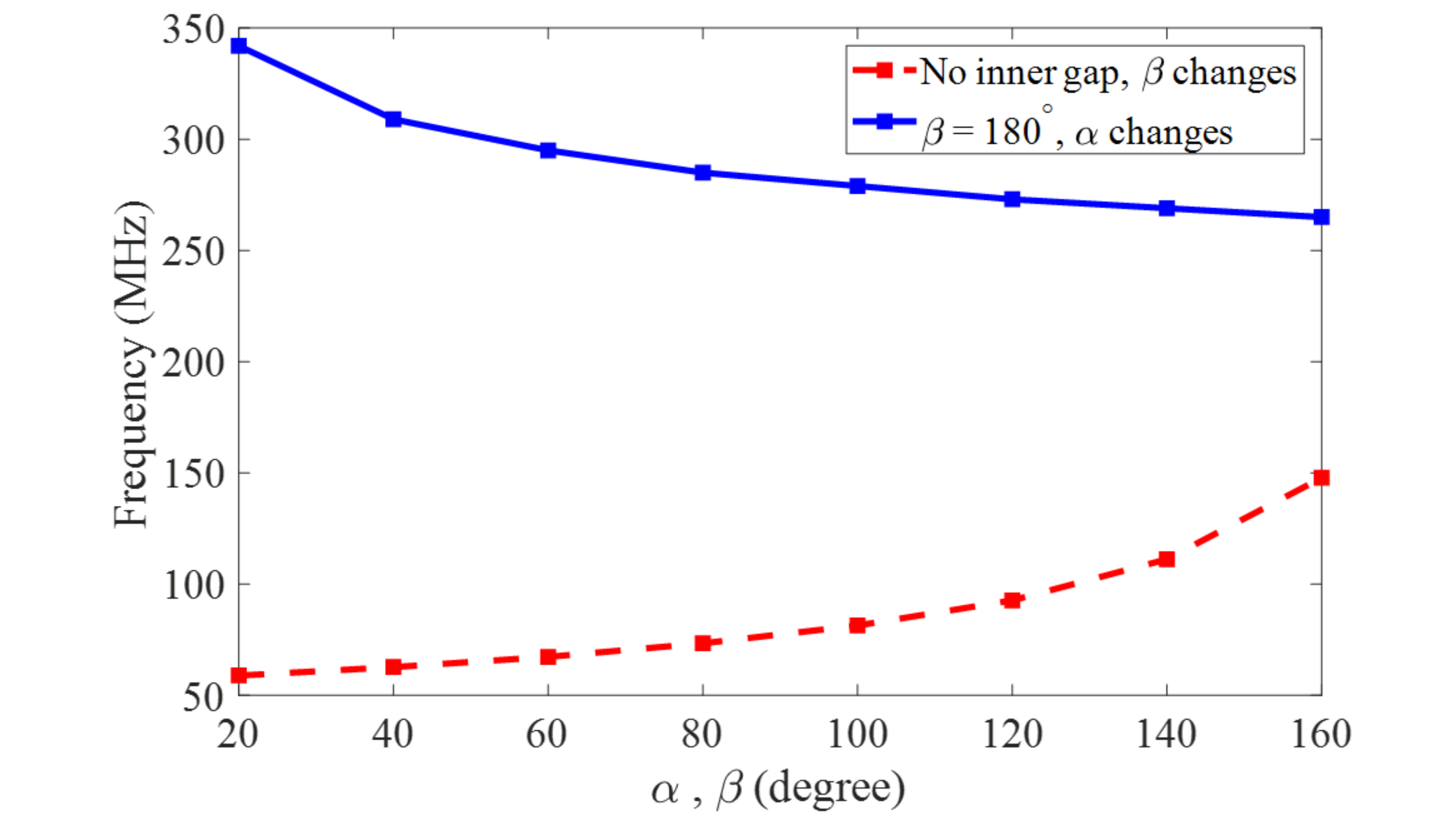}
\caption{Resonance frequency of the modified HIC with respect to the two symmetrically added gaps to the shield -- red line -- and with respect to the two symmetrically added gaps to the inner wire while there is one gap at $\beta=180^{\circ}$ on the shield -- blue line.}\label{f14}
\end{figure}

We simulated performance of our HIC using CST Microwave Studio, in both Frequency and Time Domain solvers. To verify applicability of perturbation method to our HIC, we simulated the transceiver HIC with different positions of the added perturbing gaps. In the simulation $r_0=80$mm, $b=1.2$mm, $a=0.5$mm and dielectric permittivity of the coaxial cable is $\epsilon_r=2.2$.
Similar to the HIC presented in \cite{r12}, there is a gap on top of our transceiver HIC to make the coil radiative. To raise the resonance frequency we added two symmetrically positioned cuts to the shield and varied the angle $\beta$ shown in Fig.\ref{f13}(a) from $20^{\circ}$ to $160^{\circ}$. The resonance frequency of the coil changes as depicted in Fig. \ref{f14} -- there is no local maximum on the red line because the maximum corresponds to $\beta = 180^{\circ}$. Next, we fixed the optimal $\beta = 180^{\circ}$ and added two more symmetrically positioned cuts to the inner wire at the angle $\alpha$ changing it also $20^{\circ}$ to $160^{\circ}$. The result is a blue line in Fig. \ref{f14}. Again, as much as these added gaps in the inner wire are closer to the top, where also the electric field produced by the electric dipole mode is maximal, the resonance frequency increases rapidly. When the cuts in the inner wire are close to those in the shield, their impact is low because they start to operate as a single perturbing element. 

Thus, introducing our additional cuts we may increase the resonance frequency of our enlarged HIC from 56 MHz up to nearly 350 MHz. Now, let us optimize the angles $\alpha$ and $\beta$ so that to engineer the HIC resonance 
at 298 MHz and to minimize the mutual coupling in an array. In this regime, the electric and magnetic modes of the current are balanced. We simulated two in-plane HICs separated by a gap $3$ cm ($\lambda/33$) with antenna inputs located upside down and loaded by $L_R$ (see below) in both transmission and reception regimes. In these simulations we found $\beta=180^{\circ}$ and $\alpha=32^{\circ}$ which grant the needed operation frequency and the decoupling in the transmission regime. For two transmitting HICs we got at 298 MHz $S_{12}=-4.6$ dB. As expected, this value is lower than that obtained for two HICs from \cite{r12} ($S_{12}=-3.36$ dB in accordance to Fig.~\ref{f7}). Overlapping our transceiver HICs helps to further reduce the coupling. Note that in the presence of a body phantom with high permittivity reduces the coupling between the antennas in the array. In this case, lower coupling granted by our HICs compared to conventional coils, makes the S-parameters more stable in presence of a phantom with varying sizes and permittivity.

\begin{figure}[t!]
\centering
\includegraphics[width=1\linewidth]{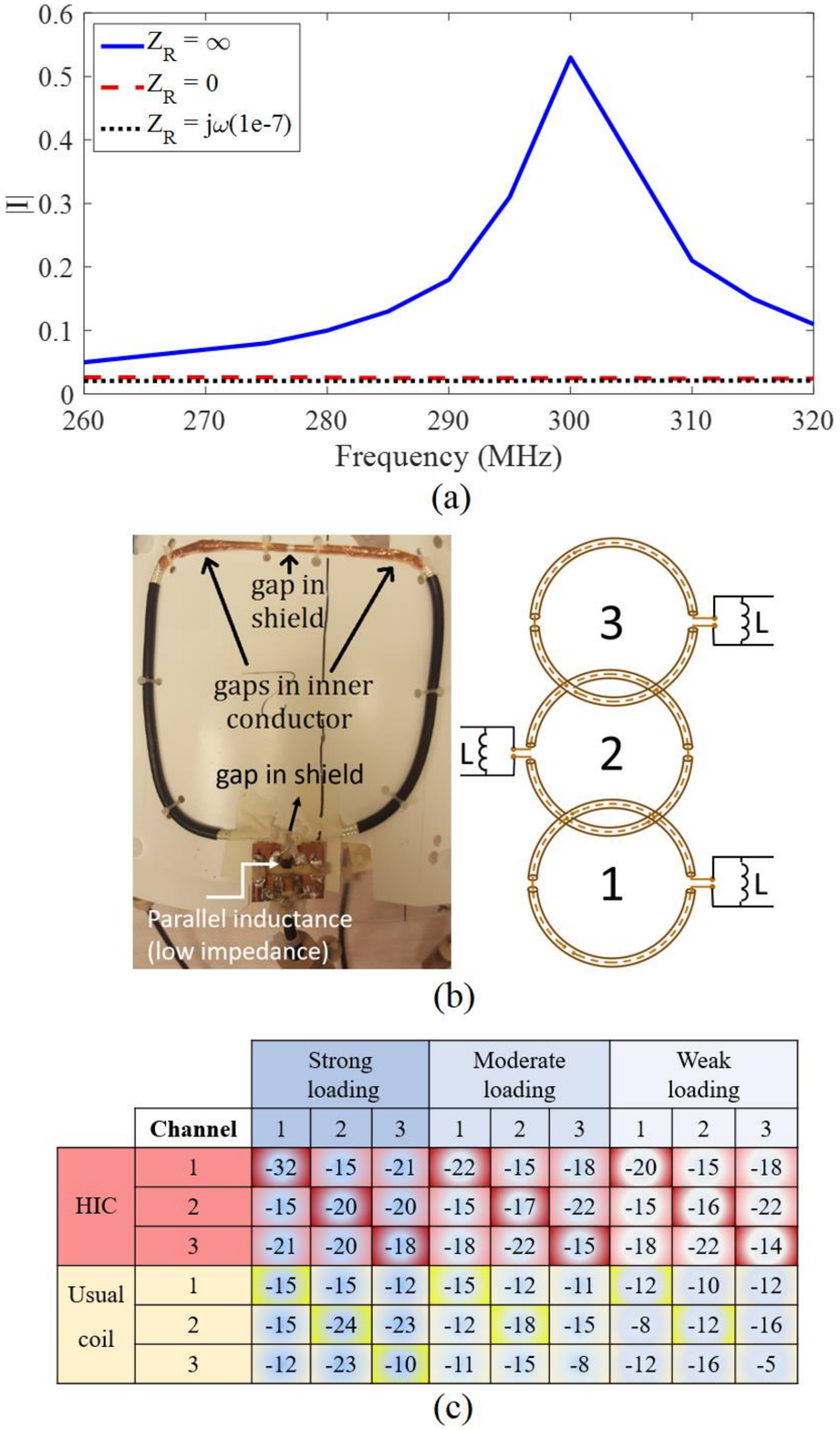}
\caption{(a) Simulation of the mean induced current in the receiving regime for three different loads at the port. (b) Fabricated transceiver HIC (with the matching circuit) and the arrangement of the ports in the array (inset). (c) Measurement results (in dB) for resonant S-parameters of the antenna array of three HICs and the same for three conventional coils in presence of three different phantoms.}
\label{f15}
\end{figure}

In the reception regime, using the optimal $L_R=100$ nH, we obtained $S_{12}$ for our HICs nearly the same as for two HICs introduced in \cite{r12}. We have also checked that for the open and shortcut receiving port the induced current at the resonance becomes large and non-uniform. Fig. \ref{f15}(a) shows the simulation result for the induced current for three values of $Z_R$. Since the induced current is not uniform over the external loop, the current in this plot is the mean value (averaged over the loop perimeter). This result confirms once more what we have claimed about the decoupling in the reception regime.

\section{Experimental Verification and Conclusions}
\label{sec:experiment}

To verify the performance of our HICs experimentally, we fabricated an array of $3$ transceiver HICs operating at 298 MHz and arranged as it is shown in Fig. \ref{f15}(b). Dimensions of our HICs were the same as those we have found above via simulations though the circular shape of the loops was slightly modified when we fixed them in the array. Fig. \ref{f15}(b) shows one fabricated transceiver HIC with its matching circuit. In the matching circuit, the inductance $L_R=170 nH$ was connected in parallel with a capacitance. This matching circuit allowed us to combine the proper $Z_R$ in the receiving regime together with the matching in the needed operation band.
Due to lower coupling between two transceiver HICs compared to conventional coils, we could match our HICs easier and the S-parameters are more stable when we vary the parameters of the phantom. Here, it is worth to note, that overlapping the loops decouples the adjacent ones quite well, but the coupling between non-adjacent loops makes S-parameters of these loops sensitive to the ambient. A slight variation of the phantom size, permittivity or conductivity causes the deviation in the resonant S-parameters and the decoupling of adjacent loops worsens.

Our HICs creating weaker fields aside should grant the lower coupling for non-neighboring elements and the array should be more stable with respect to the ambient. To verify this guess, we measured S-parameters of an array constituted of three transmitting HICs in the presence of three different cylindrical phantoms and compared the results with those for an array of three conventional capacitively loaded coils. The phantoms were as follows: (1) salty water with $\epsilon_r=75$, conductivity $\sigma=0.5$, length 200 mm and radius 75 mm (strong loading in MRI terminology), (2) salty water with reduced radius 50 mm, other parameters are same (moderate loading), (3) vegetable oil with $\epsilon_r=3$, conductivity $\sigma=0.01$, length 250 mm and radius 60 mm (weak loading). The conventional coils had the radius $r_0= 80$ mm and inserted capacitors making the loop resonant at the same frequency (298 MHz). The conventional array was optimized for the case of the strong loading. 

Fig. \ref{f15}(c) shows S-parameters of both arrays in the presence of three different loads. As we expected, both $S_{ii}$ and $S_{ij}$ of the transceiver HIC array are more stable compared to the conventional coil array. 
Our HIC really grants the better decoupling for non-adjacent coils 1 and 3. And $S_{12}$ keeps stable (-15 dB) for HICs, whereas for the loop array we have $S_{12}=-15$ dB only for the strong loading. For the moderate and weak loadings the coupling increases to $S_{12}=-12$ dB and $S_{12}=-10$ dB, respectively. Our numerical and experimental results clearly show that the suggested transceiver HIC is advantageous with respect to the conventional coils and keeps the advantages of the receiving HIC from \cite{r12} in the transmission regime.

To conclude: in this paper, physics of previously introduced high-impedance coils for MRI was comprehensively studied in both transmission and reception regimes. We have found that in the transmission regime these HICs are not advantageous. We introduced a transceiver HIC which is really beneficial with respect to  conventional coils in the range of 300 MHz. We have proved it analytically using the equivalent circuit and numerically, using CST simulations. To confirm our theory experimentally we fabricated an array of transceiver HICs and an array of conventional loops for comparison. The coupling in the transmitting regime is really lower for our HICs and our array is more stable to the variations of the human body phantom. The stability refers to both matching and decoupling.

\end{document}